\documentclass[11pt]{amsart}
\usepackage{geometry}                
\usepackage{graphicx}
\usepackage{amssymb}
\usepackage{epstopdf}
\begin{document}

\title{What do we know about the geometry of space?}
\author{B. E. Eichinger}
\maketitle
\begin{center}
{\small\itshape Department of Chemistry, University of Washington, Seattle, Washington 98195-1700}\par
\end{center}

\begin{abstract}
The belief that three dimensional space is infinite and flat in the absence of matter is a canon of physics that has been in place since the time of Newton.  The assumption that space is flat at infinity has guided several modern physical theories.  But what do we actually know to support this belief?  A simple argument, called the "Telescope Principle", asserts that all that we can know about space is bounded by observations.  Physical theories are best when they can be verified by observations, and that should also apply to the  geometry of space.  The Telescope Principle is simple to state, but it leads to very interesting insights into relativity and Yang-Mills theory via projective equivalences of their respective spaces. 
\end{abstract}

\section{Newton and the Euclidean Background}
\label{intro}
Newton asserted the existence of an Absolute Space which is infinite, three dimensional, and Euclidean.\cite{Newton}  This is a remarkable statement.  How could Newton know anything about the nature of space at infinity?  Obviously, he could not know what space is like at infinity, so what motivated this assertion (apart from Newton's desire to make space the sensorium of an infinite God)?  Perhaps it was that the geometric tools available to him at the time were restricted to the principles of Euclidean plane geometry and its extension to three dimensions, in which infinite space is inferred from the parallel postulate.  Given these limited mathematical resources, there was really no other choice than Euclid for a description of the geometry of space within which to formulate a theory of motion of material bodies.    

In this context, one of the boldest statements in all of science is Newton's First Law: "Every body continues in its state of rest, or of uniform motion in a right line, unless it is compelled to change that state by forces impressed upon it."  The first part immediately begs the first question -- with respect to which coordinate frame is a body at rest?  The second part of the statement, "continues in a state . . . of uniform motion in a right line"  is a brilliant abstraction.  One can state with some assurance that no one in the 17th century had seen uniform motion in a straight line persist for more than a brief interval of time.  Newton's abstraction enables the third part of the statement, hinting at an equation, the existence of which is established in the Second and Third Laws.  Newton's definition of motion also required the notion of Absolute Time, thereby building mechanics upon the notions of flat space and time.  

Newton's contemporaries, Berkeley\cite{Berkeley} and Leibniz\cite{Alexander}, criticized the ideas inherent in the First Law.  The gist of their objections is that motion, or lack thereof, that Newton takes as \emph{a priori} observable, cannot be defined except in relational terms. Philosophers, being literal minded, may have been trying to understand the First Law logically, as a statement about a single object.  If one imagines a single object imbedded in Newton's three dimensional Euclidean space, how can one define whether it is stationary or moving?  In the mind's eye we might imagine such an object and space -- it is a harder exercise to remove the mind's eye from the picture and contemplate nothing but the object and the void.  It now seems logically impossible to make any statement about motion of any kind, be it linear or rotational (as in the rotating bucket argument).\footnote {See also Cajori's Appendix Note 13 in Vol. II of ref. \cite{Newton} for a more thorough discussion.  The issues that are raised by the First Law have been discussed thoroughly in sources to numerous to cite here.}  On further reflection, it seems impossible to make an incontestable statement about what is meant by ``space" for this thought experiment.  Assuming the guise of a modern differential geometer, it is no more difficult, at least mathematically, to imagine a three-dimensional physical object submerged in, say, a 37-dimensional complex space than in a three-dimensional Euclidean space. 

Further thought to the definition of place and motion reveals that these are always dependent on relations.  More precisely, and within the flat geometry of Euclid, four non-coplanar points are required to define a coordinate frame: the origin and three directions defined by lines connecting the origin with the other three points.  The motion of a a fifth object or point is dependent on the choice of frame, and can only be made quantitative in terms of relations between five points!  Of course, one might now object that we want theories that are invariant to the choice of frame, but this simply equates the issue to invariance with respect to all sets of four non-coplanar points, and does not circumvent the basic problem.  It is also difficult to understand the meaning of Absolute Time.  Can observable time be defined in the absence of a sequence of distinguishable events?  Of course, all this is intended to expose the idea that motion cannot be defined except in relative terms; there is more to Newton's First Law than meets the eye.

So what?  Newton produced a successful mechanics, Leibniz and Berkeley did not.  What can be done with a theory couched exclusively in terms of the relations between objects?  Newton had the wisdom to produce a succinct set of laws with which he could compute using a very flexible coordinate system, so philosophy be damned.  And the subsequent success of that program in all areas of the quantitative sciences cannot be assailed.  

In the Scholium, Newton acknowledged the existence of relative location in space, which is a happy expedient for calculations.  Subsequently, the general answer to this coordinate frame ambiguity is provided by an inertial frame.  Frames that are not in constant rectilinear motion, such as a frame affixed to the rotating earth, may require fictitious forces to describe motions within them.  There is more than a hint of something special happening with rotational motion, to which I will return later.

In the next major development in the use of Euclidean space and time coordinates, Maxwell distilled the observed behavior of  electricity and magnetism into the laws of the electromagnetic field.  The underlying premise of electromagnetic field theory reflects back to the First Law.  Newton's \emph{body} becomes a test particle in a force field.  To make this work, the test particle does not appreciably perturb the field.  If there is a significant perturbation, that is, if the test particle and the object that creates the field are of comparable influences on one another, a self-consistent calculation is required in all but the simplest case.  If only two charges are considered, such as for the Bohr atom, the problem is solved in a frame fixed on one of the particles, and the field is derived from a symmetric potential.  However, the motion of two electrons in a field is an unsolvable classical three body problem.  Quantum mechanical extensions to many-body theory generally require an iterative self-consistent calculation, where the mathematical techniques resolve to the calculation of the motion of one particle in a field created by other particles, as in Hartree-Fock or density functional theory.  

General relativity is also a theory of the motion of a test particle in a field (created by a massive object), but now the conceptual leap made by Einstein was to identify geometrical curvature with force. Once more, the idea behind the theory is that the test particle does not perturb the field.  It should be added here that Einstein was wedded to flat space in the absence of matter; General Relativity requires space to be flat at infinity.  (This was stated differently and carefully by Einstein, where he required the theory to admit a \emph{finite} region of spacetime where special relativity holds.\cite{Einstein}   In the construction of the field equations, this domain is not finite, but appears as a boundary condition at infinity.  A moment's reflection reveals that a \emph{finite} domain anywhere else would constitute a flat spot in an otherwise curved space  -- very strange indeed.) 

In stating that an object moves in a straight line in the absence of forces, Newton defined a flat tangent plane to the 3D space of position.  Given this simple statement, one sees that, as a material object moves it carries a tangent plane along with it.  If the tangent plane tilts or rotates as the motion progresses, the reorientation is described as curvature in differential geometry, which equates curvature to force.  Now geometry supersedes force.  The simplest classical example of this effect is required in the description of the motion of an object on a rotating sphere: ``fictitious forces" are produced as the tangent plane is being reoriented by the sphere's rotation. This is a purely geometrical effect, resulting simply from the sphere being a 2D non-Euclidean manifold whose tangent plane is being reoriented by the rotation.\footnote{Reorientation of the tangent plane requires an external frame of reference to verify motion, so we return to the need for an external inertial frame as the conceptual framework for rotational problems.}  The ideas of General Relativity suggest that the difference between a ``fictitious force" and a ``real force" is inconsequential; the adjectives are based on Newtonian ideas of causality. Force is replaced by a \emph{connection form} in field theory.  

Now that we have a geometrical home for ``force", differential geometry teaches that we should be thinking of local charts and atlases to avoid making \emph{a priori} assumptions about the manifold that best describes nature.  We do not have to discard local Euclidean geometry, but instead should be prepared to acknowledge that spacetime coordinates might best be understood in terms of local charts.   In the spacetime context, a local chart is a region of the origin in ${\mathbb R}^4$.  The smaller the curvature, the more nearly will the coordinates on the local chart conform to the natural coordinates on the manifold. For example, the classical equation for the electric field $\phi,\nabla^2\phi=\delta({\bf r})$, might best be understood as an equation in a local chart. 

With this very brief historical summary one arrives at the geometrical side of today's physics.  The other side, in which forces are manifest as contact interactions between bosons and fermions, paints a very different picture.  That picture is what this work is directed at, so no further comment is made here.  However, there is an overarching philosophical statement that does need to be stated: The Newtonian philosophy contained in the First Law is simply not set up as a many-body theory, and that legacy pervades all of physics.  One can say that Newton's Euclidean space is \emph{The} foundation of physics.  Is it inevitable?

\section{Telescope Principle}
The main contention of this paper is very simple to state, but it requires a context to make it digestible. A part of this context has been discussed, and now we turn attention to another part of the argument, again invoking a pseudo-historical evolution of ideas.  

As stated in the Abstract, we should form our understanding of the nature of space on the basis of observations.  Clearly we locate objects with three suitably chosen coordinates and we measure distances between objects with use of a convention that a assigns measure based on a mapping between a number space, ${\mathbb R}^3$, and Euclidean space, ${\bf E}^3$.\cite{Boothby}  This distance measure is assigned to astronomical objects.

As astronomers became adept at measuring astronomical distances with telescopes, each advance in optical  power (or other part of the electromagnetic spectrum) resulted in an ever increasing distance scale.  Let these successive distance scales be $R_1,R_2, . . . $.  At successive times in history, corresponding to the progressive index $i$ of $R_i$, one could state that all observable phenomena occur within a sphere of, say, radius $10R_i$, to be generous.  That is, one could only claim to \emph{know} something about the relation between the observer and object if the object were no more distant than $R_i$, but to be on the safe side we'll say distances $r<10R_i$.  The word \emph{know} is in italics to signify that it is to be understood in the context of scientific thought or theory, where one assigns an objective reality to the thing being observed.  (Very much to the point is that distances measured by Doppler shifts depend on assumed Euclidean geometry.)  At any stage in the history of astronomical or cosmological observation one may scale distances by the current largest distance, and thereby arrive at $r_i=r/10R_i<1$.  To the present day, the \emph{observable} universe is limited to this three dimensional ball.\footnote{See ref. \cite{KN} for a presentation of this geometry that does not scale distance.}

Three possibilities are discussed for the geometry of the universe: positive, zero, or negative curvature.  Of these possibilities, the only background or global curvature that renders the previous statement patently invalid is positive, for which it is logically possible for a limiting distance to be reached.  This observation has not been made, so let's see what can be done with our limited range of observation.\footnote{The reader will by now have seen that most, if not all, of the conceptual and mathematical ideas that are discussed here are textbook topics, and can be found on in a variety of sources, including Wikipedia.}

\section{Projections into the Poincar\'e Ball}
The scaling, by $10R_i$, invoked in the previous section is not necessary to what we are going to do in this section, but it makes the mathematics a little less cumbersome.  The Poincar\'e ball is embedded in ${\mathbb R}^3$ and is defined by  $r<1, r=|{\bf r}|, {\bf r}\in {\mathbb R}^3$.  As seen from outside, the ball is embedded in Euclidean space, so that distances are finite; from inside the ball the intrinsic geometry is non-Euclidean, and the distance from any point to the boundary is infinite.\cite{KN}   That is, observations from within the ball cannot detect the boundary.  This is a hyperbolic space, but it is only one ``model" for such a space.  

The other two standard models of hyperbolic space are the upper-half-space (which finds use in physics, but which will not be further discussed here) and the two sheeted hyperboloid.  Turning attention to the hyperboloid, define the vector ${\bf x}$ with components $(x_1,x_2,x_3)$ in a global ${\mathbb R}^3$ frame, and the family of hyperboloids
\begin{equation}\label{hyperboloid}
x^2_0-{\bf x}{\bf x'}=\sigma^2
\end{equation}
where $\sigma\in {\mathbb R}^+$ is a real number, $x_0\in {\bf R}$, and ${\bf x'}$ is the transpose of ${\bf x}$.  This hyperboloid is mapped into the ball with ${\bf r}={\bf x}/(|x_0|+\sigma)$, which is easily seen since
\[
{\bf r}{\bf r'}={\bf x}{\bf x'}/(|x_0|+\sigma)^2=(x^2_0-\sigma^2)/(|x_0|+\sigma)^2=(|x_0|-\sigma)/(|x_0|+\sigma)\le 1
\]
From eq. (\ref{hyperboloid}) it is clear that $|x_0|\ge\sigma$, and the inequality follows.  Both  branches of the hyperboloid are handled with $x_0\to |x_0|$.  The focal point of the $x_0>0$ branch is at $-\sigma$, and at $+\sigma$ for the $x_0<0$ branch. Two coordinate charts are required to cover the two branches of the hyperboloid.

Given just the fact that observations are limited, one deduces the existence of spacetime coordinates with Lorentz signature.  But there is yet another projection that is very important, and this is from the sphere to the ball.  The three dimensional sphere is defined by 
\begin{equation}\label{sphere}
y^2_0+{\bf y}{\bf y'}=\rho^2
\end{equation}
where ${\bf y}=(y_1,y_2,y_3)\in{\mathbb R}^3$.  The projection ${\bf r}={\bf y}/(\rho+|y_0|)$ gives
\[
{\bf r}{\bf r'}={\bf y}{\bf y'}/(\rho+|y_0|)^2=(\rho-|y_0|)/(\rho+|y_0|)\le 1,
\]
and the inequality is again easy to prove.  The southern hemisphere, with $y_0<0$ is mapped to the Poincar\'e ball with focal point at the north pole, and the northern hemisphere maps to the same ball with a focal point at the south pole; the equator maps to the boundary of the ball.  (Again, two charts are required to cover the sphere.)  This gives us a mapping from the hyperboloid to the ball to the sphere, thereby mapping from relativistic spacetime to the instanton.\cite{tHooft}  Presumably a theory in one frame can be mapped into a theory in the other frame, with an important qualification to be discussed.

\section{Matrix Representations of Surfaces}
In preparation for the final section, the representation of these spaces, hyperboloid and sphere, will be switched to a matrix basis.  Taking up the hyperboloid first, eq. (\ref{hyperboloid}) can be expressed in a matrix form, with basis elements in the Pauli representation, as the determinant of 
\[
X=\left[{\begin{array}{*{20}c}
x_0+x_3& x_1+ix_2\\
x_1-ix_2& x_0-x_3\\
\end{array}}\right]=\left[{\begin{array}{*{20}c}
\chi_a& \chi_c\\
\bar\chi_c& \chi_b\\
\end{array}}\right];\quad \textrm{det}(X)=\sigma^2,
\]
As every graphics programmer knows, a three-vector $t$ in the quaternion representation, which is just $\sqrt{-1}\times$(Pauli representation), is rotated with a unit quaternion, $u\in SU(2)$, by $t\to\hat t= utu^{-1}=utu^*$, where $u^*$ is the transpose conjugate of $u$.  Conjugation by $u$ acts only on the three-vector part of $X$ in $X\to\hat X =uXu^{-1}$ and does not affect the identity component $x_0$, which commutes with $u$. To involve $x_0$ in a larger space of transformations, one has to expand the algebra: $SU(2)\to SL(2,{\mathbb C})$, for which the determinant evaluation of $X\to\tilde X=sXs^*, s\in SL(2,{\mathbb C})$, remains invariant.  This representation and group operation is the basis of the isomorphism $SO(3,1)\sim SL(2,{\mathbb C})/\pm 1$.

One may operate in the same way with the sphere.  In the matrix representation the sphere, eq. (\ref{sphere}), is 
\[
Y=\left[{\begin{array}{*{20}c}
y_0+iy_3& y_1+iy_2\\
-y_1+iy_2& y_0-iy_3\\
\end{array}}\right]=\left[{\begin{array}{*{20}c}
\zeta_a& \zeta_b\\
-\bar\zeta_b& \bar\zeta_a\\
\end{array}}\right];\quad \textrm{det}(Y)=\rho^2,
\]
with $s:Y\to sYs^*$, as for the hyperboloid, to transform the determinant and the sphere.  However, the matrix $Y$ can also be interpreted as a representation of a quaternion, which admits a real quadratic form as a product $YY^*=\rho^21$, where $1$ is the unit matrix and $Y^*$ is the transpose conjugate of $Y$.  But the determinant operation is foreign to the quaternion ring, which is a clue that we are entering a different domain with a different set of rules. 

To proceed, we need Hamilton's \emph{quaternion representation} corresponding to the matrix representation of $Y$.  The quaternion basis consists of four elements: $\{{\bf 1,i,j,k}\}$, where ${\bf 1}$ commutes with the other three elements which anti-commute amongst themselves, as can be proved from the definitions:  ${\bf i}^2={\bf j}^2={\bf k}^2={\bf ijk}=-{\bf 1}$.  A quaternion $q$ has components: $q=q_0{\bf 1}+q_1{\bf i}+q_2{\bf j}+q_3{\bf k}$ and a conjugate $q^*=q_0{\bf 1}-q_1{\bf i}-q_2{\bf j}-q_3{\bf k}$ with real coefficients, $q_\alpha, 0\le\alpha\le 3$.  The quaternion algebra is distributive but not commutative, which identifies it as a ring and not a field, but it is a division algebra.  

A sphere arose as an ``instanton" \cite{tHooft} in Yang-Mills (YM) theory.\cite{YM,Belavin} The relation between a sphere, an instanton, and a quaternion is contained in Atiyah's proof \cite{At} that the curvature two-form, $F$, that minimizes the Yang-Mills functional is 
\begin{equation}\label{Curv}
F=(1+q_pq_p^*)^{-2}dq_p\wedge dq_p^*.
\end{equation}
where $q_p$ is a quaternion.  The geometry that gives this curvature two form is a coset space, $Sp(2)/Sp(1)\times Sp(1)=Sp(2)/Sp(1)^2$ (see, for example, ref. \cite{Hua,Lawson}); it is also a projective space.  The $n$-dimensional symplectic group, $Sp(n)$, is the ``unitary" group over the quaternions:  $Sp(n):=U(n,{\mathbb H})$, defined by $g\in Sp(n), gg^*=1$, where $g^*$ is the matrix transpose of $g$ with elements that are quaternion conjugates of those in $g$. 

A matrix $x(q)$ in the coset $Sp(2)/Sp(1)^2$ is of the form 
\[
x(q)=\exp\left[{\begin{array}{*{20}c}
0& q\\
-q^*&0\\
\end{array}}\right],
\]
where $q$ is an arbitrary quaternion.  In a physical application, it is natural to think about propagating a motion as a geodesic.  Happily there exists a theorem\cite{Price,Simon} to this effect: ``geodesics are cosets of one parameter sub-groups"\cite{Sternberg}.  That is, $x(tq), t\ge 0$, is a geodesic path on the group for fixed $q$.  Factoring the magnitude $\omega = |q|=\sqrt{qq^*}$ gives $\xi=tq=t\omega u, uu^*={\bf 1}$, or $\xi\xi^*=\omega^2t^2$, which is the equation for a sphere (an expanding sphere as $t$ increases), the same as eq. (\ref{sphere}) in real coordinates.  For physical applications, $\omega t=|q|t$ is \emph{dimensionless}. For completeness, the geodesic on the coset is  
\begin{equation}\label{sp2}
x(tq)=\exp\left[{\begin{array}{*{20}c}
0& t\omega u\\
- t\omega u^*&0\\
\end{array}}\right]=\left[{\begin{array}{*{20}c}
\cos(\omega t){\bf 1}& \sin(\omega t) u\\
- \sin(\omega t) u^*&\cos(\omega t){\bf 1}\\
\end{array}}\right]
\end{equation}
where the unit quaternion $u\in S^3$, is a point on the unit three-sphere.  

It should be clear that this is more than just an exercise in rearranging the terms in the hyperboloid to get a sphere: $x^2_0-{\bf xx'}=\sigma^2\to x^2_0={\bf xx'}+\sigma^2$.   This is also not a Wick rotation: $t\to \sqrt{-1}t$.  The Yang-Mills functional yields a geometry that incorporates an \emph{instanton}, which features in both the tangent space $\mathfrak {sp}(2)$ and in the projective space\cite{Eich1} inherent in eq. (\ref{Curv}).  Group theory provides a connection between a geodesic path and the propagation of the instanton.  The projection from the hyperboloid to the ball: ${\bf x}/(c|t|+\sigma)$ and from the sphere to the ball: ${\bf y}/(\omega|t|+|x_0|)$ are very similar in their relations between space and time coordinates.  However, the way that $t$ appears in the Lie group setting is different from that in Maxwell's equations and Special Relativity, and this is very subtle issue.  I purposely wrote the coordinates on the hyperboloid without identifying $x_0$ with $ct$.  The relation between the hyperbola and sphere (instanton) establishes the time dependence of the hyperbola through a one-parameter geodesic path on the group, with the path parameter being identified with time. 

Let ${\mathfrak g}$ be an infinitesimal operator in the Lie algebra of a coset of the group $G$.  An equation of the form 
\[
d\Psi/dt= \mathfrak {g}\Psi
\]
with solution 
\begin{equation}\label{QM}
\Psi(t)=\exp(t\mathfrak {g})\Psi(0)
\end{equation}
evolves $\Psi$ along a geodesic in $G$.  These equations can be viewed as generalizations of standard quantum mechanical methods in both the Schr\"odinger and Dirac equations.  They are perfectly natural descriptions of the action of geodesics in group theory.

\section{Many-Body Theory}
There are four themes at work throughout this discussion:  (\emph{i})motion and position are defined by relations between objects, (\emph{ii})spacetime coordinates are local chart coordinates, (\emph{iii})a tangent space provides a local chart, and (\emph{iv})projectivities relate apparently different geometries to one another. This leads to the notion that a theory formulated with spacetime coordinates in a local chart might admit a mapping to a theory formulated with instanton coordinates in a Lie algebra, provided that only one particle, \emph{i.e.}, only one coordinate frame, is considered.  Going beyond the map from the hyperboloid to the sphere, the structure of the symplectic coset space provides the possibility for a new interpretation of coordinates and interactions that synthesizes a connection between eq. (\ref{QM}) and the four themes.

The action of $Sp(2)$ in its fundamental representation requires two quaternion-valued components in the vector space on which it acts.  For $g\in Sp(2)$ this action is 
\[
g\Psi=x(tq)H\Psi=\left[{\begin{array}{*{20}c}
a&b\\
c&d
\end{array}}\right]
\left[{\begin{array}{*{20}c}
\psi_1\\
\psi_2
\end{array}}\right]
\]
where $H$ is the subgroup $Sp(1)\times Sp(1)$.  Factoring the action of $H=h_1\times h_2 \in Sp(1)\times Sp(1)$, and using eq. (\ref{sp2}), gives
\[
g\Psi=\left[{\begin{array}{*{20}c}
\cos(\omega t){\bf 1}& \sin(\omega t) u\\
- \sin(\omega t) u^*&\cos(\omega t){\bf 1}\\
\end{array}}\right]
\left[{\begin{array}{*{20}c}
h_1\psi_1\\
h_2\psi_2
\end{array}}\right].
\]
But the action of $h_i$ on $\psi_i$ has no effect on the magnitude of $\psi_i$ -- it is a gauge group, identical to the isotopic gauge group of YM.\cite{YM}  In other language, the quaternion $\psi_i$ has an $Sp(1)$ fiber sitting on it.  The projection $\pi:Sp(2)\to Sp(2)/Sp(1)^2$ from the $Sp(2)$ bundle space to the base space factors the ``ineffective" part of the group from the action that couples the two components of $\Psi$ to one another.  The action of $Sp(1)\times Sp(1)$ is ineffective because it leaves each component unaffected by the other. 
[It may be useful to count real dimensions: dim$(Sp(n))=n(2n+1)$. $Sp(2)$ is 10 dimensional, 3 each in $h_1$ plus 1 in $\omega t$ and 3 in $u$.]

One may interpret the two quaternion-valued components of $\Psi$ as spins, which follows from the fact that $Sp(1)\sim SU(2)$ acts on an elementary spin.  This interpretation is consistent with the notions expressed in the Introduction, that nature insists that we can only discover relations between objects.\footnote{The Dirac equation also harbors a relation between two objects -- an electron and a coordinate frame for the free electron or an electron and nucleus for a bound electron.}  Although Yang and Mills\cite{YM} appeared to be formulating a theory for two ${\mathbb C}$-valued states, the mathematics took control and insisted that there be a relation between two ${\mathbb C}^2$-valued objects.  These objects are understood to be elementary fermions, and the contact between them is provided by a boson that resides in the group.  Near the identity of the group, $q$ acts on $\psi_2$ while simultaneously $-q^*$ acts on $\psi_1$.  

In summary, the instanton structure of Yang-Mills theory supplies a picture in which fermions reside in a representation space, and the interaction between them is mediated by a boson residing in the coset space $Sp(2)/Sp(1)\times Sp(1)$.  Given the mapping from the sphere to the hyperbola, we no longer have any other contact with Euclidean space -- the only coordinates available to us are contained in the group and its algebra.  In particular, the instanton must not be imbedded in Euclidean space.\cite{tHooft}

This structure immediately suggests an extension to $Sp(3)/Sp(1)^3$ to describe the structure of three fermions.  The group $Sp(3)$ contains an $SU(3)$ subgroup, which promises further insight into the structure of quark composites.  Three quarks in a proton, presumably all moving relative to one another, might be better described in this interaction picture than in an inertial frame fixed to the stars -- what matters are the relations between the quarks -- not an external observer's frame.  The advantage of the quaternion algebra is that representations of composite particles, constructed from sums and differences of components in the tensor product of the vectors in the representation space, remain in the quaternion ring.  This is not the case with Dirac spinors.   Work on the irreducible representations of $SU(3)$ is underway.

\section{Conclusion}
This discussion of the legacy of Newtonian Euclidean space went through an acknowledgement that relations between objects are implicit in physical theories, and led up to a discussion of a projective equivalence between relativistic hyperbolic geometry and the spherical geometry inherent in Yang-Mills theory.  The geometry that is inherent in YM theory led to the interpretation of the role of $Sp(2)/Sp(1)^2$ as the action of a boson on a pair of fermions.  This picture immediately suggests an extension to the flag manifold $Sp(3)/Sp(1)^3$ to represent the space of interactions between three fermions.  There is yet a further extension to a very large family of flag manifolds, $Sp(n)/Sp(k_1)\times Sp(k_2)\times\cdots\times Sp(k_m); \Sigma^m_ik_i=n$, as a new description of the symmetries of nature.  Interactions between subsets of particles in this picture are encapsulated in curvature two-forms.\cite{Eich2} The action of the coset space, represented as a flag manifold, on its representation space is a self-consistent many-body action.  Multiplication of an element in the representation space by an element in the group is a contact interaction between a boson and a fermion.  Simultaneously, another element in the representation space is acted on by the negative conjugate boson.  This ``simultaneous through space"  interaction between the two fermions changes the meaning of causality at the level of elementary particles.

\end{document}